# Inductively Coupled Plasma at Atmospheric Pressure, a Challenge for Miniature Devices

**Horia-Eugen, Porteanu[1], Ilija Stefanović[2,3], Michael Klute[4], Ralf-Peter Brinkmann[4], Peter Awakowicz[2], Wolfgang Heinrich[1]**

*[1]Ferdinand-Braun-Institut, Leibniz-Institut für Höchstfrequenztechnik, Berlin, Germany*
*[2]Ruhr-Universität Bochum, Faculty of Electrical Engineering and Information Technology, Institute for Electrical Engineering and Plasma Technology, Germany*
*[3]Serbian Academy for Sciences and Arts, Institute for Technical Sciences, Belgrade, Serbia*
*[4]Ruhr-Universität Bochum, Faculty of Electrical Engineering and Information Technology, Institute for Theoretical Electrical Engineering, Germany*

## *Abstract*

Plasma jets belong to the category remote plasma. This means that the discharge conditions and the chemical effect on samples can be tuned separately, this being a big advantage compared to standard low-pressure reactors. The inductive coupling brings the advantage of a pure and dense plasma. The microwave excitation allows furthermore miniaturization and generation of low temperature plasmas. The present paper shows the state of the art of the research on such sources, demonstrating their work up to atmospheric pressure.

## I. SOURCE AND MEASUREMENT SETUP

There are several attempts to create resonant structures, in which a strong magnetic field is designated to drive inductively a plasma [1-3]. Our option was to design a structure that concentrates the magnetic field mostly in the plasma by minimizing unuseful currents and stray magnetic fields. Pairing two sources into a double ICP was the improved solution [4,5]. The source consists of a copper block, as shown in Fig. 1, where the hole plays the role of a coil and the slit that of a common capacitor. A 7 mm outer diameter quartz tube is used to confine the plasma.

Not every inductive configuration leads to an inductive coupling for every input power. The transition from E (or CCP) mode to H (or ICP) mode is largely treated in the literature especially for sources at 13.56 MHz.

Our paper [6] makes a mode analysis using microwave measurements. Here few details are recalled in order to understand the challenges for an atmospheric pressure ICP source. In order to measure the system impedance while plasma is excited, we overlap two signals from a generator (hp8350B) and from a network vector analyzer (Rohde & Schwarz ZVA 8), Fig. 1.

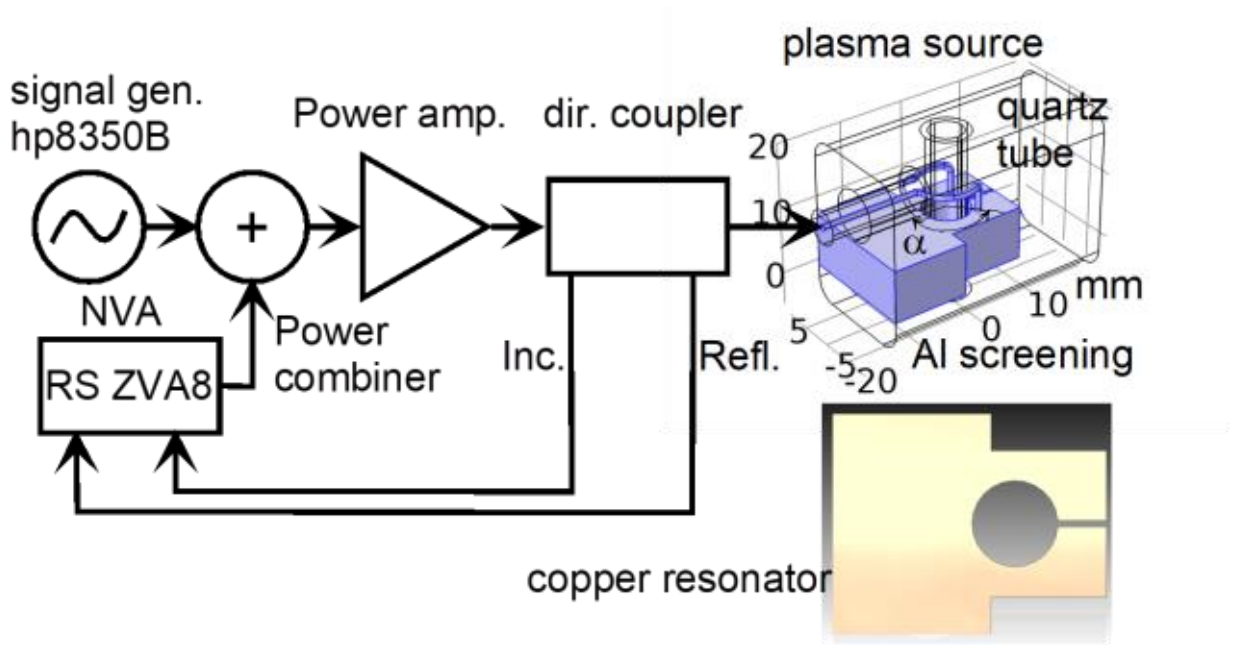

**Figure 1.** Source design and measurement setup for "Hot-S" parameters

The strong signal excites the plasma and the weak signal probes the dynamic conductivity of the plasma in a wide frequency range for a given plasma state (electron density). Briefly, the raw data are a number of resonance curves recorded for different excitation powers (Fig. 2).

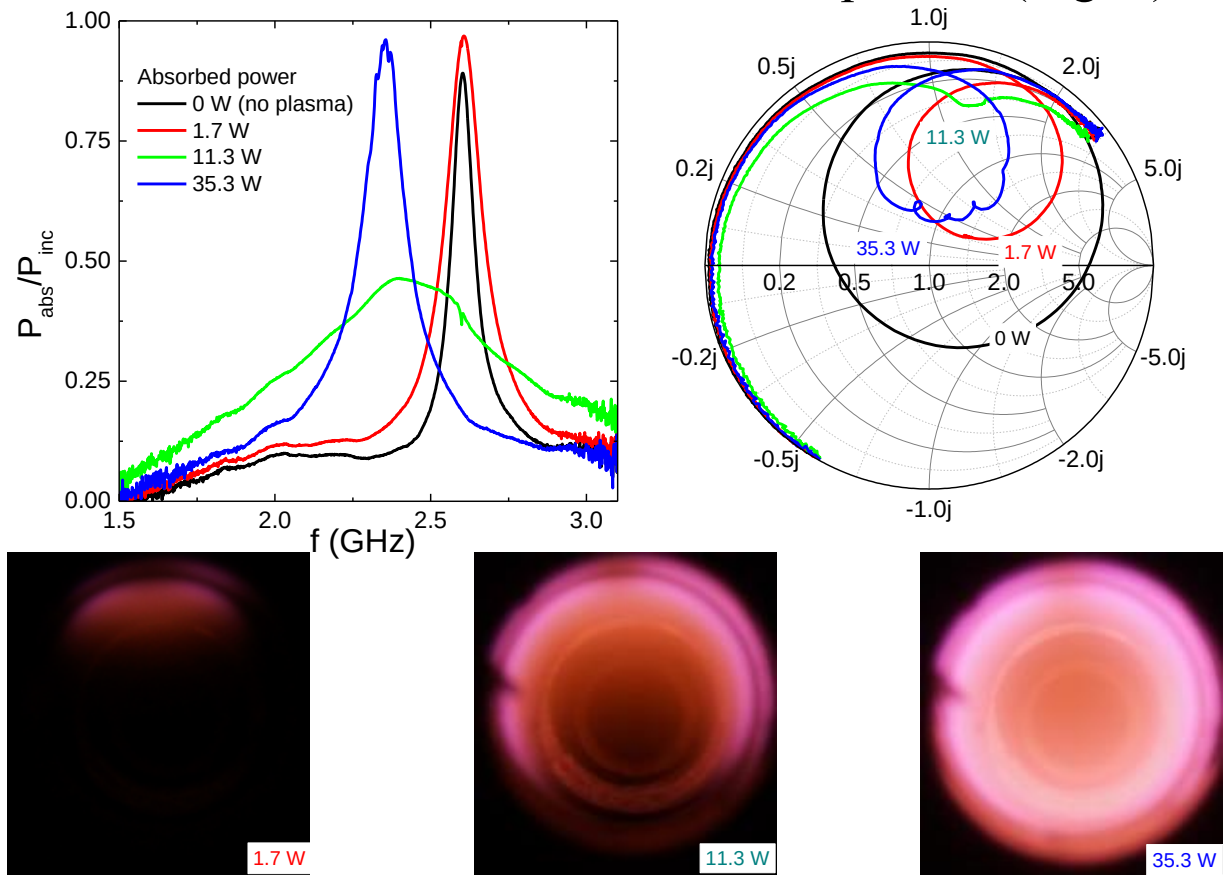

**Figure 2.** Raw data from the "Hot-S"-spectroscopy for nitrogen at 1000 Pa (top). The curves show in frequency domain (left) and in Smith diagram (right) the evolution of the relative absorbed power and of the relative impedance to 50 Ω, when the power is increased. Correlated are presented pictures of the plasma column at powers of 1.7 W, 11.3 W and 36.3 W (from left to right).

Fig. 3 left show the evolution of the resonance frequency when the excitation power is increased.

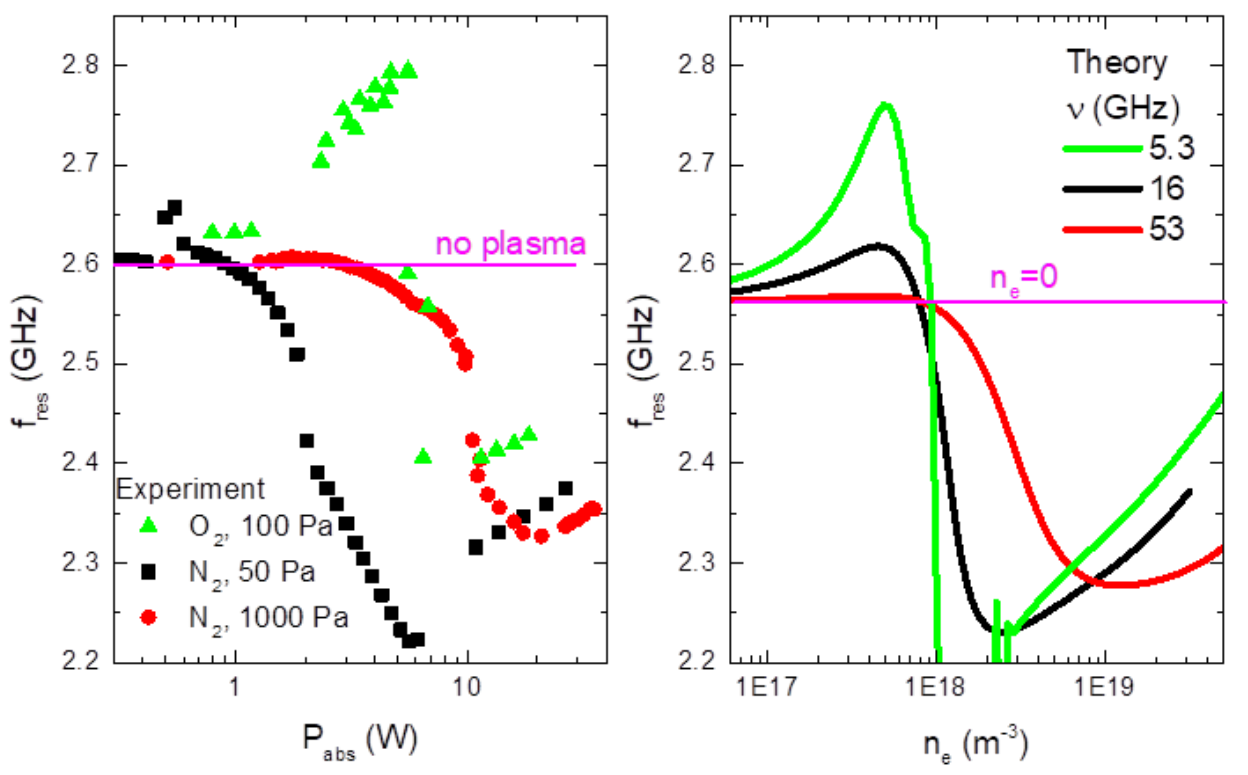

**Figure 3.** Experimentally determined variation of the resonance frequency of the source for different absorbed powers in the system (left) compared with the theoretical prediction of this variation for different electron densities (right).

# II. GLOBAL MODEL AND DATA INTERPRETATION

The inductive coupling has the advantage of a thin sheath existing near the quartz wall, which can be neglected in a first instance in the simulation. We used the cold plasma conductivity defined by the electron density and the scattering frequency. Moreover, in our model the plasma is considered homogeneous in the whole volume. Using COMSOL as a 3D simulator for the geometry presented in Fig. 1 right, we get various information like field and current distribution or the dependence of the resonance frequency on the electron density (Fig. 3 right). In spite of a very simplified model we observe a close resemblance between experimental data and theory.

The correspondence between the two curves allows us to find a dependence between the electron density in the plasma and the absorbed power (Fig. 4, left). We carried out experiments mainly with nitrogen at different pressures (100 Pa, 1000 Pa, gas flow 150 sccm) and argon at 1 atm. In order to confirm the determined electron density, optical measurements on the emission spectrum of nitrogen have been performed [7]. The optical measurements allow also the estimation of the gas temperature (Fig. 4 right).

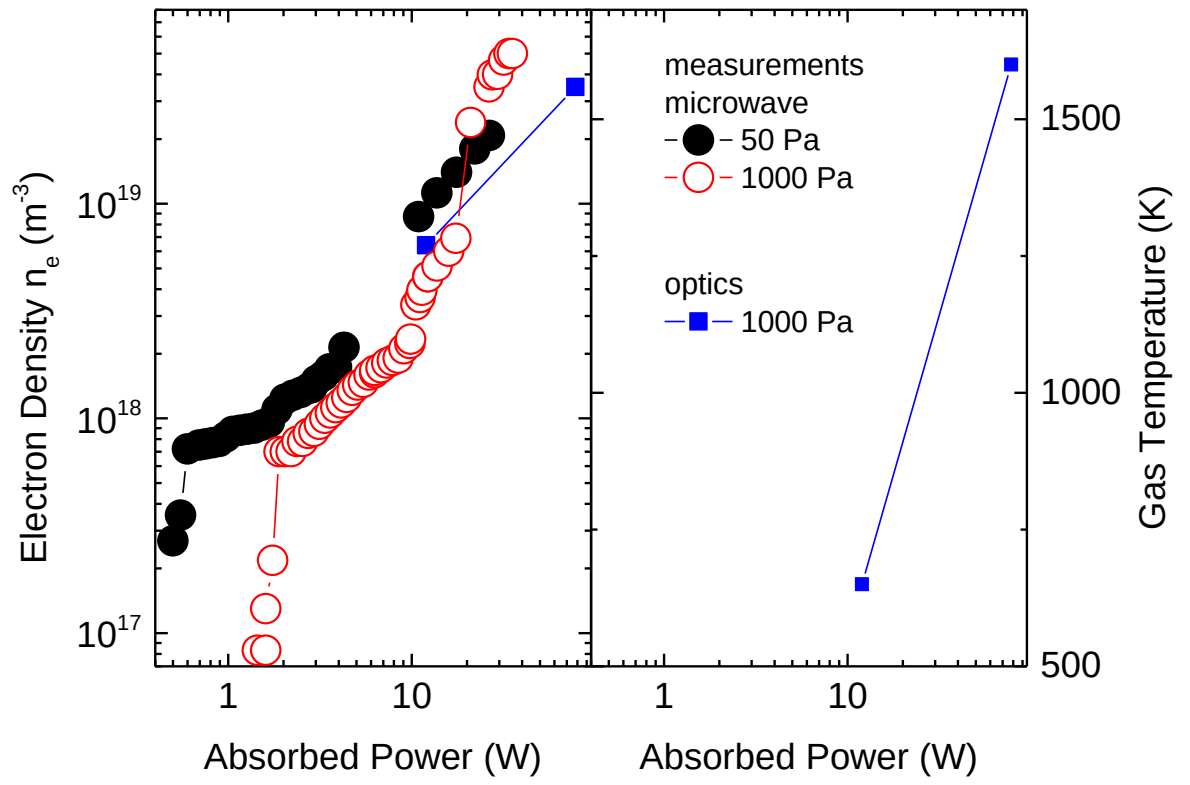

**Figure 4.** Electron density (left)as a function of absorbed power determined with our method [6] and with optical method [7] (left). Also from [7] gas temperature as a function of absorbed power (right).

The power consumed in order to reach the ICP mode at low pressure is of the order of few tens of watts. The microwave measurements allow not only to find a resonance frequency but also an impedance of the system at the resonance. If we consider the absorption in the copper block quasiconstant and mode independent (a detailed discussion is to be found in [6]), one can estimate the plasma resistance as real part of the plasma impedance at the resonance. This resistance is plotted for two pressures in Fig. 5 left. Knowing the absorbed power and using a simple lumped model of the system, consisting of a parallel resonant circuit, with the copper resistance defined by the width of the resonance without plasma, one can estimate the absorbed power in the plasma. The essential result is an absorbed power of 25 W in the plasma for 35 W incident power at 1000 Pa. The value of more than 60 % of absorbed power in the plasma is, to our knowledge, a record value.

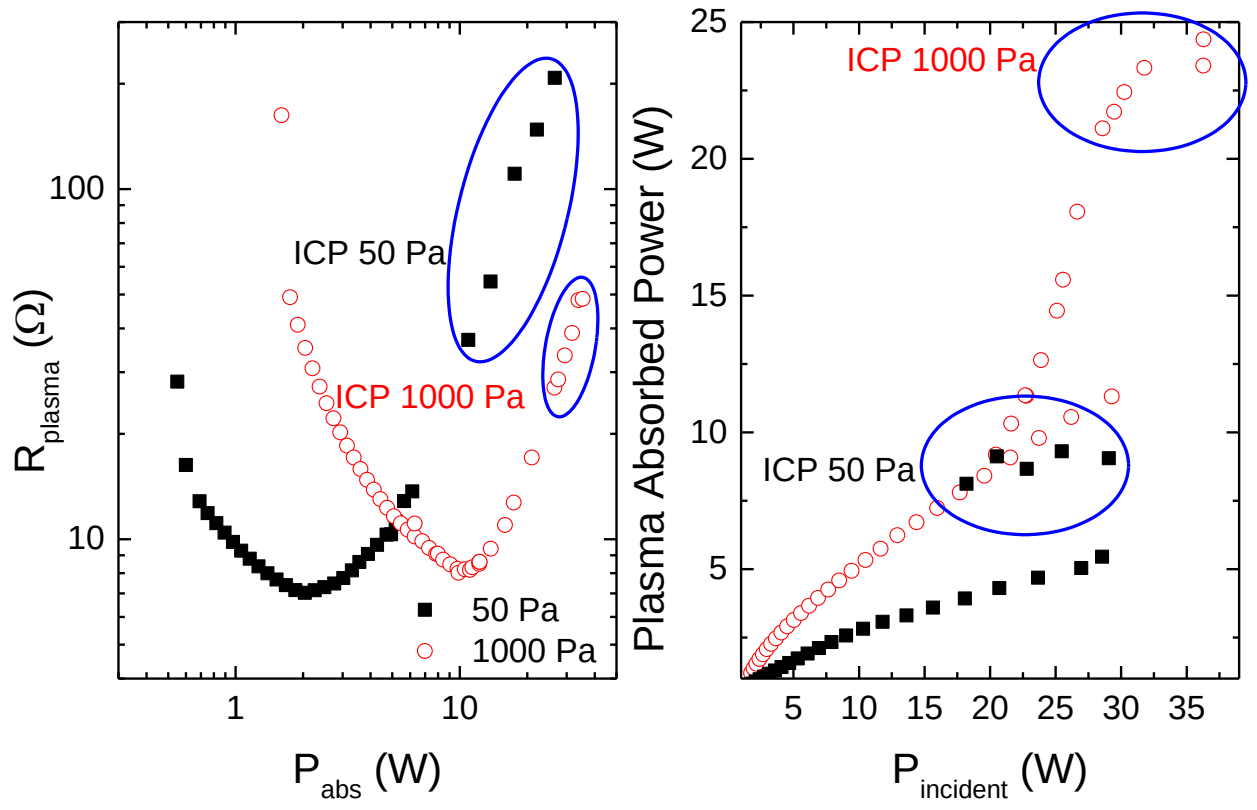

**Figure 5.** Plasma resistance as a function of absorbed power (left) and plasma absorbed power as a function of incident power (right).

## III. TESTS AT ATMOSPHERIC PRESSURE

The source has been tested with different tube diameters. While for applications a large diameter is beneficial, the number of undesired microwave modes becomes larger. In order to reach the ICP mode it is necessary to use higher power. Fig. 6 shows the behavior of an argon plasma at atmospheric pressure using an excitation power of 130 W. At lower powers the plasma luminescing ring is incomplete. Because of the high necessary power for the excitation a Kuhne generator KUSG2.45-250A, working between 2.4 and 2.5 GHz with powers up to 250 W was used instead of the scheme presented in Fig. 1.

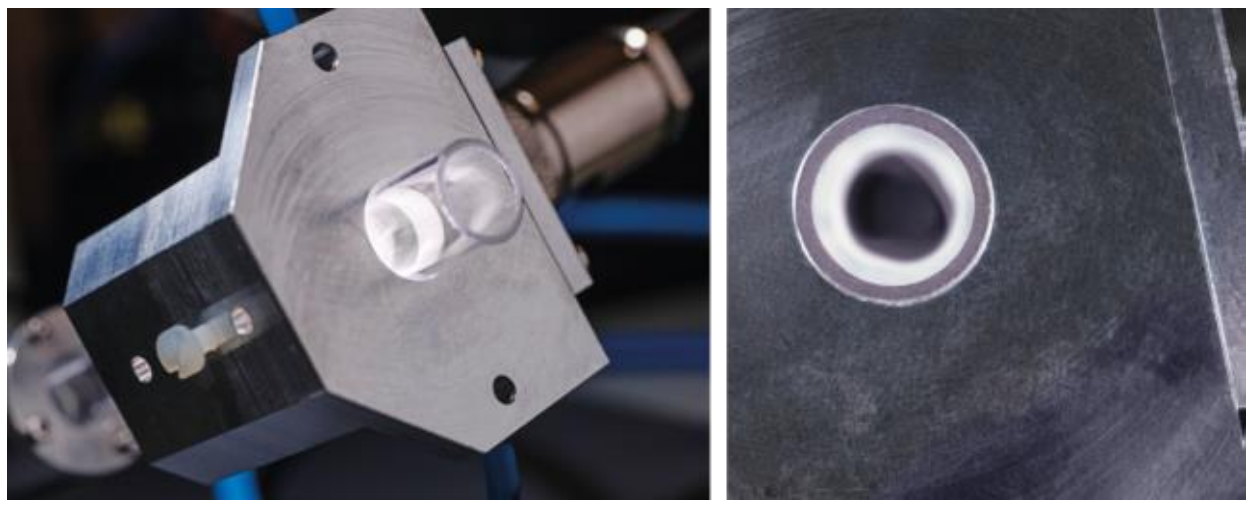

**Figure 6.** Microwave ICP source driven at atmospheric pressure with argon. P=130 W.

The impedance matching circuit is briefly presented in the inset of Fig. 1. The angle “alpha” defines the position of the connecting wires on the coil (hole) and represents the transformation ratio for the system impedance. Just the use at atmospheric pressure reveals new details related to this simple wire connection.
The exact wire position, the distance to the aluminum screening and other details can cause undesired additional CCP modes. This was the case for the source in Fig. 6. The matching system, the magnetic and electric field produced by the feeding wires and the losses here are to be analyzed in future.

Due to the diffusion characteristics the plasma has different shapes at different pressures (Fig. 7).

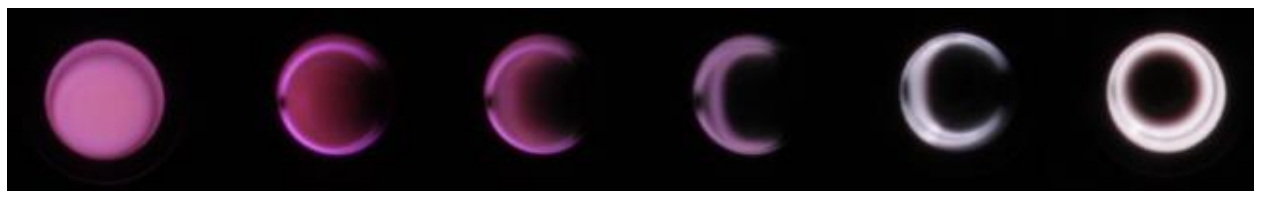

**Figure 7.** Evolution of the argon plasma shape while the pressure is increased. From left to right are presented different excitation conditions: 50 Pa, ICP, 40 W; 200 Pa, CCP, 40 W; 1000 Pa, CCP, 40 W; 5000 Pa, CCP, 40 W; 1 atm, CCP, 80 W; 1 atm, ICP, 130 W.

A further application is the use of an array of ICP sources at atmospheric pressure, Fig. 8. In this case we used two double ICP structures, as presented in [4,5]. In order to keep the experiment simple, we used two generators KUSG2.45-250A connected each to a double ICP resonator. The impedance matching is realized here with microstrip lines. A power of 160 W (2 x 80 W) is enough in order to reach an ICP mode in each tube.

The ignition generally has been performed with a 33 kHz, 10 kV generator. High voltage rings are placed outside the quartz tube. At low pressures it couples capacitively over the quartz tube wall (1 mm thickness), plasma and screening (ground). At atmospheric pressure we placed temporarily ignition electrodes inside the tubes. At this pressure the lack of ignition of one or more tubes is not a big disadvantage. It does not create a strong disbalance in the impedances such that a further ignition would not be possible, On the contrary, the tubes can be ignited safely step by step if simultaneous ignition does not occur. The array is designed in such a way that an extension to a large number of equidistant jets in both x and y directions is possible.

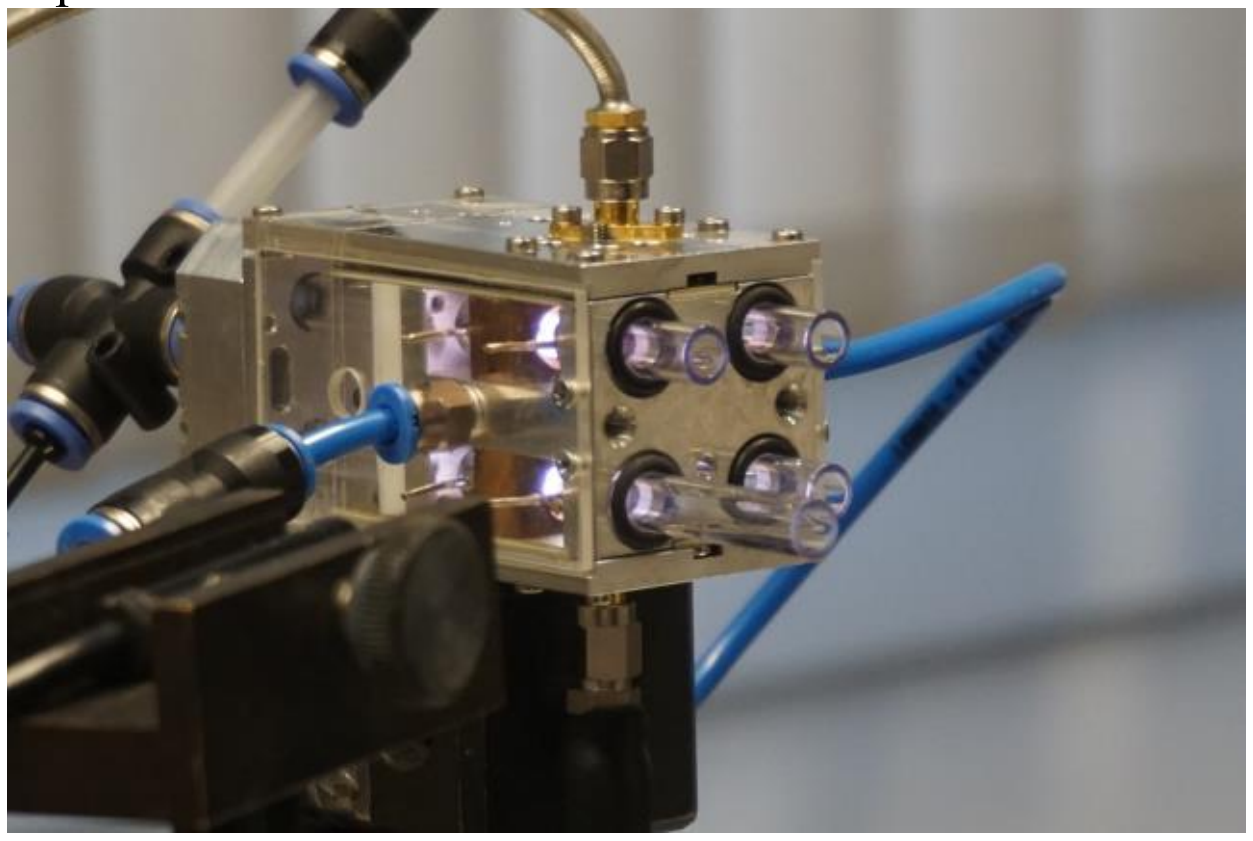

**Figure 8.** Array of four ICP sources driven with microwaves (2 x 80 W) and argon.

The use of microwave ICP sources at atmospheric pressure opens a large number of potential of applications, among them deposition, etching, activation, up to medical treatments. Where are these sources better than the existing CCP sources? A higher density of electrons and of reactive species is always an advantage. While in the CCP case denser plasma leads to lower impedance and to bad matching, in the case of ICP this impedance increases (Fig. 5) allowing an efficient energy transfer just at high electron densities. Especially in atomic spectroscopy, if ppm traces of species are to be identified, impurities caused by the interaction with electrodes or even with dielectric barriers diminish the sensitivity of the instrument.

The challenges, however, in the development of such miniature sources are to create tunable resonators with variable matching impedance in order to use a large number of gases or mixtures of them.

## II. SUMMARY

The present work answers positively the question whether an ICP plasma jet, using our original resonator design, can work at atmospheric pressures. With 130 W one can drive an argon ICP plasma in a 12 mm diameter source. An array of four 7 mm sources can be driven with argon with 160 W.

Actual microwave measurements using the "Hot-S"-Parameter spectroscopy are in good agreement with the theory and show the evolution of the electron density with the absorbed power. The scattering frequency is (mainly) a pressure dependent parameter and is kept constant for a theoretical electron density scan. With the above assumption one can determine also this parameter, considering the strength of the frequency deviation as a function of absorbed power. The determined values (Fig. 3) between 5 and 50 GHz for different pressures and atom species are in good agreement with the literature data. Further analyses and optimizations, especially for the atmospheric pressure use are planned.

## III. ACKNOWLEDGMENT

This project is supported by DFG, project No. 389090373